# Decoherence imaging of spin ensembles using a scanning single-electron spin in diamond

*Lan Luan[1]†, Michael S. Grinolds[1], Sungkun Hong [1,2], Patrick Maletinsky[3], Ronald L. Walsworth[4] and Amir Yacoby[1] ***

1. Department of Physics, Harvard University, Cambridge, Massachusetts 02138, USA;

2. Vienna Center for Quantum Science and Technology (VCQ), Faculty of Physics, University of Vienna, A-1090 Vienna, Austria;

3. Department of Physics, University of Basel, Klingelbergstrasse 82, Basel CH-4056, Switzerland;

4. Harvard-Smithsonian Center for Astrophysics, Cambridge, Massachusetts 02138, USA;

The nitrogen-vacancy (NV) defect center in diamond has demonstrated great capability for nanoscale magnetic sensing and imaging for both static and periodically modulated target fields. However, it remains a challenge to detect and image randomly fluctuating magnetic fields. Recent theoretical and numerical works have outlined detection schemes that exploit changes in decoherence of the detector spin as a sensitive measure for fluctuating fields. Here we experimentally monitor the decoherence of a scanning NV center in order to image the fluctuating magnetic fields from paramagnetic impurities on an underlying diamond surface. We detect a signal corresponding to roughly 800 $\mu_B$ in 2 s of integration time, without any control on the target spins, and obtain magnetic-field spectral information using dynamical decoupling techniques. The extracted spatial and temporal properties of the surface paramagnetic impurities provide insight to prolonging the coherence of near-surface qubits for quantum information and metrology applications.

The NV center in diamond provides convenient spin-state optical initialization and readout[1], long spin coherence times[2], and controllable proximity near the diamond surface[3], all in a robust solid under ambient conditions. It has been successfully applied to sensitive nanoscale measurements of magnetic fields[4-8] at the level of a single electron spin[9] or a small ensemble of nuclear spins[10,11]. These measurements typically employ periodic modulation of the target magnetic field in synchronization with control of the sensor NV spin[4]. However, such modulation is not possible when studying fluctuating magnetic fields found in many biological and physical systems. A method for detecting fluctuating fields that monitors the changes in coherence of the NV spin has been theoretically proposed[12-14], numerically simulated[15], and experimentally implemented to study the surrounding spin bath in the same piece of the diamond[16-19]. In particular, efforts to study external spins so far have focused on detecting diffusing spins in solution[20]. Here we image fluctuating magnetic fields from spin ensembles on a diamond surface by monitoring changes in the coherence of a separate scanning NV spin. Taking advantage of the scanning capability, we control the distance between the NV magnetometer and the sample, from which we obtain the spatial and temporal properties of the sample fluctuating fields.

The fluctuating magnetic fields originating from electron surface states are thought to strongly reduce the coherence of near-surface quantum systems. There is broad interest in characterizing such surface states as they play a key role in the demanding challenge of placing a quantum system close to a surface without compromising its coherence. This is regarded as one of the most important technical problems in a range of systems, including superconducting qubits[21], donor spins in silicon[22], and NV centers in diamond[3,23,24]. For example, to achieve high sensitivity in magnetic sensing using NV centers, it is desirable to engineer NV spins within a few nanometers to the surface[25]. However, very shallow NV spins have coherence times of only a few microseconds[3], much shorter than those deeply embedded in the bulk[2] due to electron spin ensembles on the diamond surface[23,24]. To further understand the issue of surface-induced decoherence for NV spins, we image the fluctuating magnetic fields from a nearby diamond surface and obtain the density and fluctuation spectrum of the electron spin ensembles.

We perform decoherence imaging at room temperature using a home-built scanning system that combines a confocal and an atomic force microscope (AFM) as described in ref. 26. The sensor (fabrication details discussed in ref. 27) is a single NV center artificially created through ion implantation residing few tens of nanometers from the end of a diamond scanning nanopillar 200 nm in diameter (Fig. 1a). The NV spin state is optically initialized and read out via spin-dependent fluorescence[1], and controlled by on-resonant microwave (MW) magnetic fields. The sample we study is a separate piece of single crystalline diamond, on which we pattern mesas that are 50 nm in height and 200 nm in diameter and leave the surface oxygen-terminated after the fabrication process (detailed in sample section).

Prior to imaging the fluctuating magnetic fields from electron spin ensembles on the diamond surface, we obtain the "bare" coherence time $T_2$ of the scanning NV spin by running a Hahn-echo sequence with the scanning tip fully retracted from (and thus unaffected by) the sample (Fig. 1c). We observe a decay of the NV spin coherence with evolution time $t$, in addition to coherence oscillations induced by coherent coupling to nearby $^{13}C$ nuclear spins [28]. We fit the curve with a stretched exponential envelope $\text{Exp}(-(t/T_2)^p)$ multiplied by a periodic Gaussian function $\sum_i \text{Exp}(-(t - i\tau_R^{echo})^2/\tau_{dev}^2)$, where the revival period $\tau_R^{echo} = 16.7~\mu s$ is determined by the $^{13}$C nuclear Larmor frequency in the applied DC magnetic field and $\tau_{dev}$ is the revival peak width. From the fit we extract $p = 1.7$ and $T_2 = 79~\mu s$. In contrast, $T_2$ is strongly reduced to roughly 15 $\mu s$ when we bring the nanopillar into contact with the sample. This reduction of $T_2$ occurs when the nanopillar-to-sample vertical separation is smaller than 15 nm, suggesting that the increased decoherence emerges from the fluctuating fields on the sample surface, which is strongly suppressed with increasing tip-sample distance.

We find that we are able to prolong the NV spin coherence in contact with the sample by using dynamical decoupling (DD) sequences. As shown in Fig. 2, the measured $T_2$ scales with the number of π pulses $n$ as $T_2 \propto n^{0.72}$ when XY4, XY8, and 64–π- pulses XY8 sequences[29-31] are applied. (The extraction of $T_2$ is discussed in supplementary information.) This scaling suggests that the additional decoherence source has slower dynamics than the bare $T_2$, consistent with it originating from a surface spin bath dipolar coupled[32] to the scanning NV center.

To further study NV spin decoherence induced by the surface spins, we scan the AFM tip in contact with the sample across a mesa 50 nm in height and 200 nm in diameter. The mesa is designed to induce changes in the NV-spin-sample distance when the AFM is running in contact mode. As the scanning nanopillar follows the topography and climbs onto the mesa, the NV spin, which is roughly centered on the nanopillar axis, is lifted to the height of the mesa and thus suspended 50 nm above the surface off the mesa (illustrated in Fig. 3a). The mesa topology thereby provides a controlled way to vary the distance between the NV center and the surface electron spins for a detailed distance-dependent study of the surface-induced decoherence.

Fig. 3a shows a two-dimensional image of $P(m_s = 0)$, the probability of the NV spin being in $m_s = 0$ after a Hahn-echo evolution of $t = 40$ $\mu s$, where each pixel is averaged for 2.1 s through multiples scans. We observe a halo structure in the variation of the NV spin decoherence with NV-spin-sample distance. The dark background (marked as region 1 in Fig. 3a) and the dark central circle (region 3) correspond to strong decoherence when the NV-spin-sample distance is minimized; the bright ring (region 2) corresponds to strong reduction of decoherence when the NV is suspended on the ridge, i.e., lifted about 50 nm above from the sample surface. We observe spatially homogenous decoherence, within our detection sensitivity, in regions 1 and 3, indicating that the surface spins have uniform density to within the imaging sensitivity of our experiment.

To study the distance dependence of NV spin decoherence in more detail, we record a line scan cutting through the middle of the field of view in Fig. 3a, with a signal acquisition time of 40 s per pixel. As shown in Fig. 3b, we obtain a sharp onset of decoherence as the tip scans across the sample mesa: $P(m_s = 0)$ drops to 0.5 (indicating lost coherence) with a transition width of 35 nm (90% change).

Spectral properties of the surface electron spins can be extracted from such NV-spin decoherence images. The coherence of the NV spin is lost over time due to the magnetic environment following the general expression[33]

$$C(t) = \exp(-\chi(t)) = \exp(-\frac{1}{\pi}\int_0^\infty d\omega S(\omega) \frac{F_t(\omega)}{\omega^2}) \tag{1}$$

where $S(\omega)$ is the spectral function of the fluctuating magnetic fields and $F_t(\omega)$ is a filter function corresponding to the resonant MW pulse sequences applied on the NV spin, with $F_t(\omega) = 8\sin^4(\omega t/4)$ for a Hahn-echo sequence[34].

In the present experiment, NV spin decoherence arises primarily from two surface electron spin baths: on the scanning nanopillar and on the sample. We model both spin baths as a fixed density of randomly oriented electron spins with uncorrelated flip-flops. In this case, the spectrum of each spin bath can be characterized by a Lorentzian with two parameters[33,34]

$$S_{t,s}(\omega) = \frac{\Delta_{t,s}^2}{\pi} \frac{\tau_{C_{t,s}}}{1+(\omega\tau_{C_{t,s}})^2} \qquad (2)$$

where $\tau_C$ is the correlation time among spins in the bath, and $\Delta$ is the average coupling strength of the spin bath to the NV center, which depends on the distance to the NV center as well as the spin density on the surface. The subscripts $t$ and $s$ refer to the nanopillar spin bath and sample spin bath respectively.

When the nanopillar is retracted away from the sample, the NV center is not affected by the sample spins. Thus we obtain the nanopillar spin bath parameters $\Delta_t$ and $\tau_{C_t}$ by fitting the out-of-contact spin-echo measurement using equations (1), (2) and $S(\omega) = S_t(\omega)$. Fig. 1b shows the fit with resulting values: $\Delta_t = 0.059 \pm 0.002$ MHz, and $\tau_{C_t} = 14.4 \pm 3.4$ μs.

When the nanopillar approaches the sample surface, $\Delta_s$ increases and NV spin decoherence induced by spins on the sample surface becomes significant. To estimate $\Delta_s$, we sum over the magnetic dipole coupling between and the NV center and individual electron spins on the sample surface:

$$\Delta_s^2 = \frac{1}{\hbar^2} \sum_i \left(\frac{\mu_0}{4\pi}\right)^2 \frac{1}{r_i^6} \left(\frac{3(\boldsymbol{m_0} \cdot \boldsymbol{r_i})(\boldsymbol{m_i} \cdot \boldsymbol{r_i})}{r_i^2} - \boldsymbol{m_i} \cdot \boldsymbol{m_0}\right)^2 \tag{3}$$

where $\boldsymbol{m_0} = \hbar\gamma(sin\theta\hat{x} + mcos\theta\hat{z})$ is the magnetic moment of the NV, $\theta \approx 54.7^0$ is the angle of the NV axis to the surface normal direction, $\hbar$ is the Planck constant, $\gamma = 2\pi \times 28$ GHz/T is the electron gyromagnetic ratio, and $\boldsymbol{r_i}$ is the displacement from the $i$th surface spin to the NV spin. We assume the surface spins have the same gyromagnetic ratio as the NV spin so that the magnetic moment of the surface spin $\boldsymbol{m_i} = \boldsymbol{m_0}$. For a flat infinite surface with uniform spin density $\sigma_A$, we obtain:

$$\Delta_s^2 = \frac{1}{\hbar^2} \iint dA \left(\frac{\mu_0}{4\pi}\right)^2 \sigma_A \frac{1}{r^6} \left(\frac{3(\boldsymbol{m_0} \cdot \boldsymbol{r})^2}{r^2} - \boldsymbol{m_0}^2\right)^2 = \frac{3\mu_0^2 \hbar^2 \gamma^4 \sigma_A}{128\pi d^4} \tag{4}$$

where $d$ is the distance between the NV spin and the sample surface.

We fit the in-contact echo measurement in Fig. 1b using equations (1) and (2) with $S(\omega) = S_t(\omega) + S_s(\omega)$, $\Delta_t$ and $\tau_{C_t}$ as determined above, and $\Delta_s$ and $\tau_{C_s}$ as fit parameters. We obtain $\Delta_s = 0.31 \pm 0.04$ MHz and $\tau_{C_s} = 6.8 \pm 6.2$ μs (70% confidence intervals). We note that $\Delta_s \gg \Delta_t$, suggesting higher spin density on the sample surface than on the tip surface. This is consistent with the additional cleaning and scraping of the nanopillar surface that we performed right before every measurement, which may remove any water or dust layer that accumulates over time in ambient conditions.

A precise determination of $\tau_{C_s}$ and $d$ can be obtained by fitting the distance dependence of the measured NV spin decoherence , e.g., as illustrated for one line-cut in Fig. 3b. In the fit we numerically calculate the decoherence using equation (1)-(3) as a function of the nanopillar position $x$ over a 2 μm × 2 μm area and a grid size of (4 nm)$^2$ with $d$ and $\tau_{C_S}$ as fit parameters. In the calculation we account for the decoherence originating from the nanopillar spin bath as previously determined, and for the sample spin bath we assume the same $\sigma_A$ both on and off the mesa as inferred from the in-contact echo measurement. We obtain $d = 29.7 \pm 5.8$ nm and $\tau_{C_S} = 2.5 \pm 1.0$ μs (95% confidence intervals). We note that $d$ agrees well with the value deduced from other measurements on the same device[9]. We obtain $\sigma_A = 0.28$ μ$_B$/nm$^2$ from the above value of $d$ and $\Delta_s$. Both $\tau_{C_S}$ and $\sigma_A$ are consistent with other measurements on diamond surface spins[3,23,24].

As a further consistency check for our approach, we apply the model described above to calculate the NV spin decoherence for the tip in contact with the sample and the NV spin interrogated with DD sequences, using equation (1) and (2) and filter function $F_t(\omega) = 8 \sin^4(\omega t/4n) \sin^2(\omega t/2) / \cos^4(\omega t/2n)$ for $n$-$\pi$-pulse XY sequences[34]. The resulting good agreement between the calculation and the data, as shown in Fig. 2b, supports our conclusions about the properties of the diamond sample's surface electron spin bath, as extracted from the NV decoherence images.

The extended NV spin coherence under application of DD sequences suggests that decoherence magnetometry, as demonstrated here, is a complimentary technique to NV-based AC magnetometry[4-10]. Both techniques can be applied using the same NV-diamond magnetometer to study both controllable and randomly fluctuating fields. Moreover, the electron surface spins on the scanning nanopillar will not affect AC magnetometry as long as the NV spin coherence can be prolonged by DD sequences.

In the measurements presented and analyzed here, we detect about $\sigma_A \cdot \pi d^2 \approx 790 \mu_B$ surface paramagnetic spins on the diamond sample, with SNR = 4.0 for 2.1 s integration time. The sensitivity can be greatly improved with shallower NV centers in the nanopillar as indicated by equation (4), where $\Delta^2 \propto \frac{1}{d^4}$. For example, at $d = 15$ nm we estimate a sensitivity of 50 $\mu_B$ with the same SNR. The spatial resolution can also be improved by reducing $d$. In future work, NV-diamond decoherence magnetometry and our high-resolution scanning imaging technique could be applied to studying the fluctuating magnetic fields from interesting nanoscale biological and physical systems such as ion channels in cell membranes[14], quantum fluctuations in spin liquids[35], and quantized spin-carrying surface states in topological insulators[36,37]. Decoherence magnetometry may also be implemented in other atomic-scale two-level systems such as phosphorous donors in Si[38].

**Sample**

The sample we study is a (100)-oriented single-crystalline electronic-grade diamond. Isolated mesas 50 nm high and 200 nm in diameter were created by reactive ion etching with ebeam lithography defined masks. Before measurement the sample was cleaned in boiling acid mixture of 1:1:1 sulfuric, nitric and perchloric acid and then thoroughly rinsed in distilled water. This procedure removes residues from the fabrication and leaves an oxygen-terminated surface.

AUTHOR INFORMATION

* Address: 11 Oxford Street, LISE 610, Cambridge, Massachusetts, 02138, USA

Email: yacoby@physics.harvard.edu. Phone: 617-495-1180. Fax: 617-495-4951

†Present address: Department of Physics, University of Texas at Austin, Austin, Texas, 78712, USA.

ACKNOWLEDGMENTS

The authors thank Element Six for providing diamond samples for both the sensor and targets. M.S.G. is supported through fellowships from the Department of Defense (NDSEG programme) and the National Science Foundation. S.H. acknowledges support from the Kwanjeong Scholarship Foundation when at Harvard University. This work was supported by the DARPA QuEST and QuASAR programmes and the MURI QuISM.

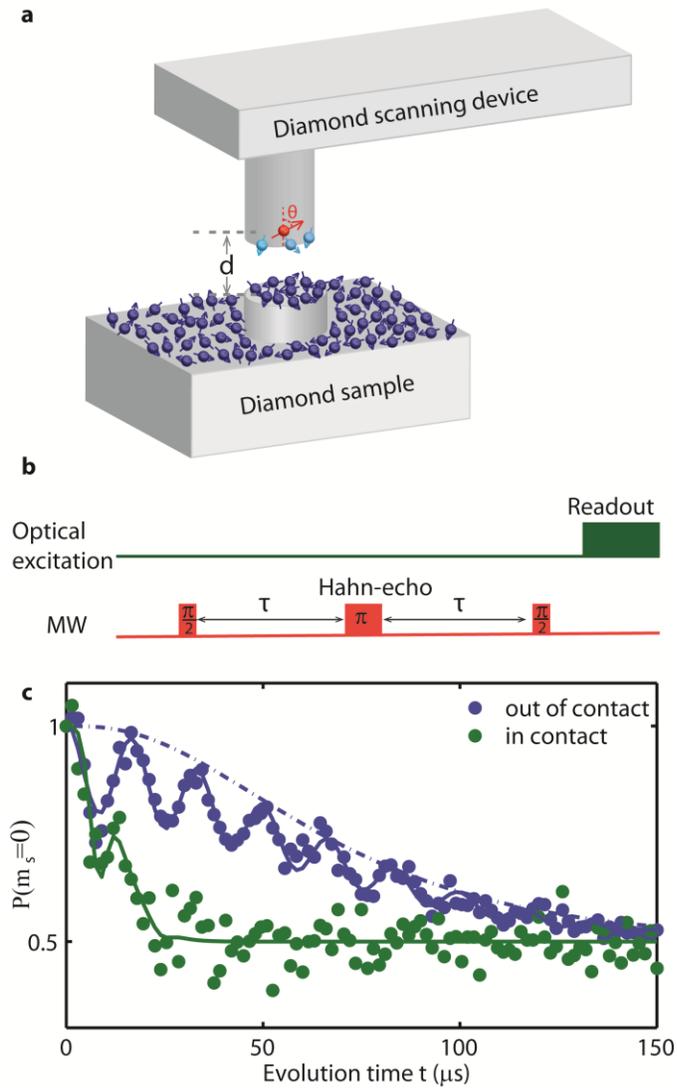

**Figure 1.** Detecting surface electron spin ensembles using a scanning magnetometer composed of a single nitrogen-vacancy (NV) spin. **a**: Schematic of the experiment. The NV spin resides in a nanopillar fabricated on a diamond scanning platform, with distance $d$ from the sample surface. The angle between the normal direction of the surface and the quantization axis of the NV spin $\theta$ is imposed by crystallographic directions. The NV spin is optically initialized and read out from above. A nearby antenna is used to apply microwave pulses to manipulate the NV spin state coherently. The diamond sample contains a mesa that is 200 nm in diameter and 50 nm in height, with paramagnetic

impurities on the surface. The NV spin is decohered by both the surface spins on the nanopillar and on the sample. **b:** Hahn-echo experimental protocol (with evolution time $t = 2\tau$). **c:** Measurements of the NV spin state ($m_s = 0$) in the scanning nanopillar after application of the experimental protocol in **b**, showing strong reduction of NV spin coherence when the nanopillar is in contact with the sample compared with it retracted 4 μm from the surface of the sample[27]. The blue solid line shows the stretched exponential fit with Gaussian shaped peaks for the collapses and revivals from nearby $^{13}$C nuclear spins. The blue dotted dash line and the green solid line show the fits to the spin bath model as described in the text.

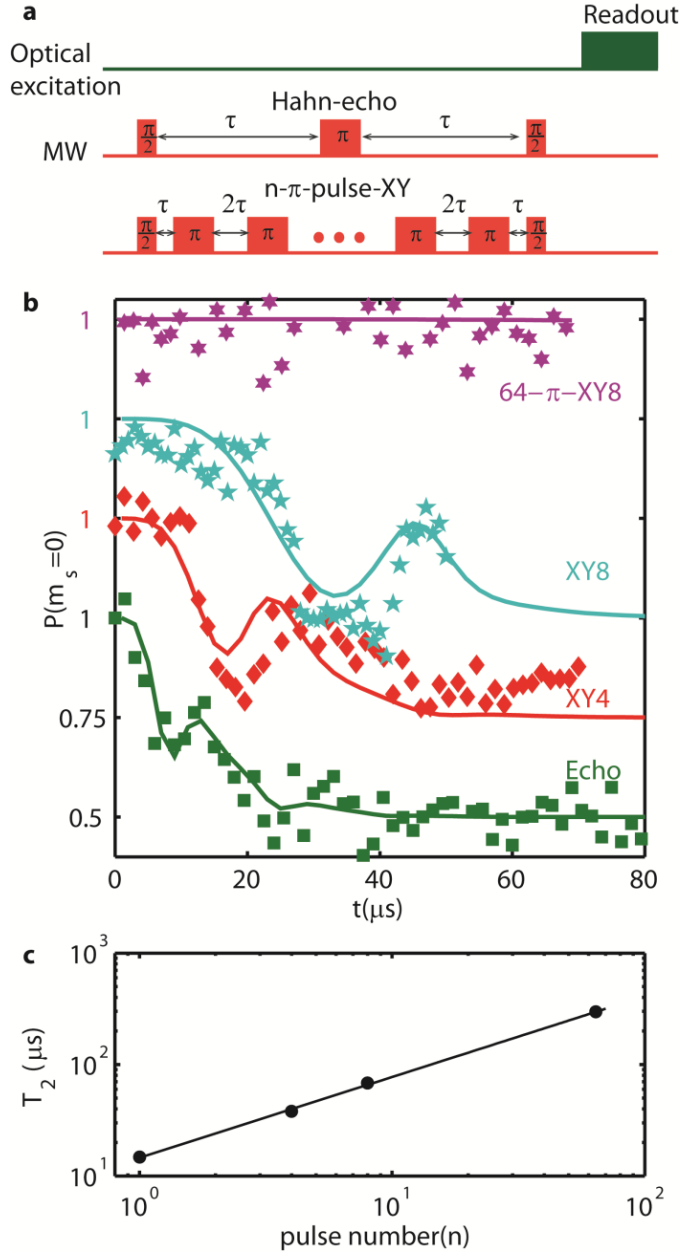

**Figure 2**. Measured and modeled coherence of the NV spin under dynamical decoupling (DD) sequences when the scanning nanopillar is in contact with the sample. **a:** Optical and MW pulses used in DD measurements with total evolution time $t = 2n\tau$, where $n$ is the number of $\pi$-pulses in the Hahn-echo and XY sequences. **b:** Measurements of the NV spin state as a function of $t$ running DD sequences as labeled. The curves labeled as XY4, XY8 and 64-$\pi$-XY8 are shifted

vertically by 0.25, 0.5, and 0.75, respectively, for clarity. The solid lines show the calculated coherence using the model and parameters described in the main text. In the calculation we account for revivals induced by $^{13}$C nuclear spins by the Gaussian function and parameters as applied in Fig. 1c. **c:** Coherence time $T_2$ determined from measurements as a function of $n$. The solid line shows the power-law fit $T_2 \propto n^{0.72}$.

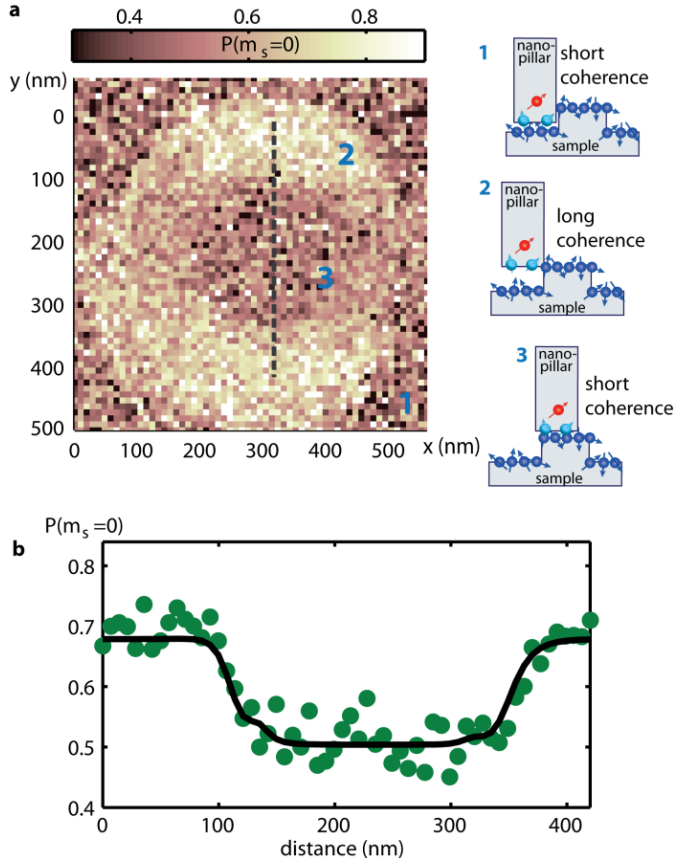

**Figure 3**. NV spin decoherence images indicating effect of surface paramagnetic impurities. **a:** Two-dimensional decoherence image as the nanopillar scans over the sample mesa while running the Hahn-echo sequence with $t = 40$ μs averaging 2.1 s per pixel. The measured coherence diminishes when the NV spin is in close proximity to the sample surface, giving the dark inner

circle and the dark outer region in the image. Also shown are sketches illustrating the relative position of the NV spin to the sample at different regions as marked in the image. **b:** One-dimensional scan along the dashed line in **a** averaging 40 s per pixel while running the Hahn-echo sequence with $t = 40$ µs. The scanning nanopillar is in contact with the sample mesa for the entire scan. The NV spin coherence is high when the NV center is suspended above the surface and has a sharp transition to zero ($P(m_s = 0) = 0.5$) when it reaches the mesa. The solid line is the fit described in the main text, from which we obtain $d = 29.7 \pm 5.8$ nm and $\tau_c = 2.5 \pm 1.0$ µs. In the fit, the NV spin is set to align with the sample mesa edges at $x = 120$ nm and $x = 320$ nm.